\documentclass[10pt]{iopart}

\usepackage{bm} 
\usepackage{graphicx} 
\usepackage[hypcap=false]{caption} 
\usepackage{amssymb} 

\usepackage{graphbox}
\usepackage{hyperref} 
\hypersetup{
	colorlinks,
	citecolor=black,
	filecolor=black,
	linkcolor=black,
	urlcolor=black
}

\begin{document}

	\title[Klein tunnelling and Hartman effect in graphene junctions with proximity exchange field]{Klein tunnelling and Hartman effect in graphene junctions with proximity exchange field}
    	\author{J. Tepper and J. Barna\'s}
	\address{Faculty of Physics, Adam Mickiewicz University, ul. Umultowska 85, 61-614 Pozna\'n, Poland}
	\ead{jedrzej.tepper@gmail.com}
	
	\begin{abstract}
		
Tunnelling of electrons in graphene-based junctions is studied theoretically. Graphene is assumed to be deposited either directly on a ferromagnetic insulator or on a few atomic layers of boron nitride which separate graphene from a metallic ferromagnetic substrate. Such junctions can be formed by appropriate external gating of the corresponding system. To describe low-energy electronic states near the Dirac points, certain effective Hamiltonians available in the relevant literature are used. These Hamiltonians include staggered potential and exchange interaction due to ferromagnetic substrates. Tunnelling in the systems under consideration is then spin-dependent. The main focus is on Klein tunnelling and also on the group delay and the associated  Hartman effect. The impact of a gap induced in the spectrum at the Dirac points on tunnelling is analysed in detail.
	
	\end{abstract}
		
	
	\submitto{\TDM}

	\ioptwocol
	
	\section{Introduction}\label{Introduction}

\indent \indent Associating wave function with material objects had several profound implications. One of the striking features was quantum tunnelling, or as it was referred to in the past penetration/leaking through the barrier \cite{Merzbacher2002}. Importantly, quantum tunnelling allowed new explanations of experimental data to emerge, with a notable historical example being the $\alpha$-decay theory. Conclusive validation of the electrons tunnelling in solids, brought by the work of Esaki \cite{Esaki1974}, marked the beginning of nowadays widespread and fruitful exploitation of this phenomenon in the area of electronics.

A question arose early on whether tunneling processes induce a delay in the propagation of the corresponding particles \cite{MacColl1932}. One of the prominent works on this problem was the paper by Hartman \cite{Hartman1962}. His analysis of a Gaussian wave packet predicted that the group delay for sufficiently thick barriers is smaller than the traversal time in the free space. Further investigations, both theoretical and experimental, led to numerous reports of superluminal tunnelling velocities. However, it wasn't until Winful's works starting from 2002 \cite{Winful2002} and culminating in 2006 \cite{Winful2006}, that the confusion was resolved and difference between the delay time and propagation delay was thoroughly explained. Currently, saturation of the group delay for thick tunnelling barriers is often referred to as the Hartman effect. This phenomenon was extensively studied in the case of particles described by Schr\"{o}dinger equation. However, it appears also in the case of other (quasi)particles, including photons \cite{Wang2004}, spin waves \cite{Klos2018}, and electrons described by Dirac model \cite{Wu2009}.

Early studies of Dirac equation provided surprising results, namely the absence of exponential damping for fermions tunnelling through potential barriers larger than the rest energy of incident particles. This phenomenon was accounted for by Klein \cite{Klein1929}, and nowadays such a property is usually referred to as the Klein tunnelling \cite{Calogeracos1999}. The problem was however of academic interest, as the potentials needed to validate the predictions of the theory lied well above capabilities of experimental physics. This was until successful micromechanical exfoliation of free standing graphene \cite{Novoselov2004} and subsequent realization that such a material could be used to test predictions of quantum electrodynamics due to similarities between two dimensional (2D) Dirac equation for massless fermions and low energy effective Hamiltonian for graphene quasiparticles \cite{Katsnelson2006}. Experimental evidence supporting theoretical predictions was presented in \cite{Huard2007,Gorbachev2008,Young2009}. After exfoliation of graphene, interest in 2D materials increased enormously and many other monolayers of atoms were obtained. Transport of electrons in junctions made of such materials, often formed by appropriate gating that allows creating a potential barrier between two outer parts of the system, was extensively studied afterwards.

A weak intrinsic spin-orbit coupling in graphene opens a gap in the spectrum at the Dirac points, which however is too small to be accessible experimentally. On the other hand, such a gap is important not only from the fundamental point of view, but also from application reasons. Such a material could be used to build a successor of metal-oxide semiconductor field-effect transistor (MOSFET) and therefore become the building block of next generation electronic devices \cite{Schwierz2010}. Thus, opening a sizeable gap in graphene band structure is a matter of ongoing scientific efforts. Yet most of the theoretically proposed and experimentally utilized methods failed to provide a sufficient gap between the valence and conduction bands in graphene or diminished desirable and unique properties of carbon monolayer such as high charge carrier mobility \cite{Xu2018}. One of the most promising methods of band structure modification is based on tailoring properties of various materials by stacking several monolayers of different substances in a specific order. While a gap in graphene separated
from nickel substrate by hexagonal boron nitride (referred to in the following as BN) has been detected experimentally \cite{Kawasaki2002}, it was only very recently when Van der Waals heterostructures received sufficient interest \cite{Geim2013}.
	
Band structure calculations based on first principles techniques proved to be useful for our understanding of properties exhibited by Van der Waals heterostructures, as well as turned out to be a valuable design tool \cite{Du2016,Qian2015}. Results obtained from such computation methods can be used to fit effective Hamiltonians, often governing quasiparticle properties in the vicinity of certain high symmetry points. Such an approach not only simplifies calculation of various phenomena, but also allows development of physical intuition often obscured by details of complicated numerical  techniques.

Recently, spin-related phenomena in graphene-based materials are of great interest. Indeed, going beyond classical logic devices based on charge carriers and exploiting spin degree of freedom in such systems emerges as a promising route of exploration \cite{Han2014}. Significant proximity-induced effects in spin-dependent band structure of graphene-based materials were predicted by several ab-initio studies \cite{Yang2013,Hallal2017,Zollner2016}. Some of the considered structures have already been successfully obtained \cite{Kawasaki2002,Swartz2012} and exchange coupling effects have been experimentally confirmed \cite{Wang2015,Leutenantsmeyer2016,Wei2016} and theoretically studied \cite{Dyrdal2017}.

In this paper we focus on graphene-based tunnel junctions, and especially on the impact of substrate-induced exchange field on tunnelling characteristics. One kind of the considered materials is graphene on nickel or cobalt with a few atomic planes of BN in between. Another group of materials to be analysed is graphene deposited directly on a ferromagnetic insulator. We address both the Klein tunnelling and Hartman effect in such heterostructures. A rectangular potential barrier is assumed as it allows detailed analysis of the fundamental characteristics and also facilitates their understanding. We analyse in detail the influence of the band gap induced by the staggered potential and exchange interaction of graphene with a ferromagnetic substrate on the Klein tunnelling and Hartman effect.
	
Section 2.1 starts with an analysis of graphene low-energy effective Hamiltonian, extended in order to encompass a larger number of 2D materials and heterostructures. Sections 2.2 and 2.3 describe a theoretical background, including calculation of tunnelling amplitudes and probabilities for a rectangular barrier, as well as wave packet dynamics and definition of bidirectional group delay. Section 3 contains numerical results for graphene on cobalt and nickel with BN in between. Section 4 contains similar analysis for graphene deposited on a ferromagnetic insulator. Summarized results together with conclusions can be found in Section 5.
	
	\section{Theoretical description}\label{Methods}
	
	\subsection{Model}\label{Model}
	
\indent \indent Linear dispersion of low-energy excitations in theoretical studies of carbon-based materials was invoked as early as in 1947 \cite{Wallace1947}. Such a dispersion emerges in tight-binding nearest-neighbour approximation for 2D hexagonal Bravais lattices with two identical atoms per unit cell (or equivalently two intersecting triangular sublattices with primitive cell). Since no intervalley interactions are considered, the corresponding  Hamiltonian can be written for individual valleys separately,
	\begin{equation}
		\mathcal{H}_{0}= \frac{\sqrt{3} a t}{2} (\tau k_x \bm{\sigma_x} + k_y \bm{\sigma_y})\otimes\bm{s_0},
		\label{h0}
	\end{equation}
where $k_x$ and $k_y$ are components of the 2D wavevector, $\bm{\sigma_i}$ ($i=x,y,z$)  are Pauli matrices acting in the sublattice (pseudospin) space, $\tau=1$$(-1)$ for the $\bm{K}$$(\bm{K}')$ valley, $t$ denotes the nearest neighbour hopping parameter, $a$ is the lattice constant, and $\bm{s_0}$ is the $2\times 2$ identity matrix in the spin space.
The prefactor in equation \eref{h0} is related to the  Fermi velocity, $\sqrt{3}at/2 =\hbar v_f$. Within this model $v_f$ is the constant quasiparticle velocity. Spectrum of the above Hamiltonian is spin-degenerate and linear with vanishing density of states at the Dirac points.
	
For pristine graphene, which exhibits the $D_{6h}$ point group symmetry, the only additional interaction in the low-energy regime is the intrinsic spin-orbit coupling \cite{Kochan2017}, so the total Hamiltonian can be written as
	\begin{equation}
		\mathcal{H}_{D6h}= \mathcal{H}_{0} + \tau\lambda\bm{\sigma_z}\otimes\bm{s_z}\equiv\mathcal{H}_{0} + \mathcal{H}_{so}^{int},
		\label{hd6h}
	\end{equation}
where  $\lambda$ is the corresponding spin-orbit coupling parameter and $\bm{s_i}$ ($i=x,y,z)$ are the Pauli matrices acting in the spin space. Note,
$\mathcal{H}_{so}^{int}$ has opposite signs for the two inequivalent valleys. The spin-orbit term opens a gap of width $2\lambda$ at both Dirac points in the electronic band structure. This gap is independent of spin.
\Eref{hd6h} can be used to fit density functional theory (DFT) calculations  near the $\bm{K}$ and $\bm{K'}$ points not only for graphene but also for other 2D materials like for instance stanene  \cite{Jiang2017} or X-hydride/halide (X=N–Bi) monolayers \cite{Liu2014}.
	
The energy gap in the low-energy spectrum can be also induced (or modified when it exists due to intrinsic spin-orbit coupling) by other interactions, especially those due to the substrate on which graphene is deposited. One kind of such interactions is the so-called staggered potential, which corresponds to inequivalent on-site energies in the two sublattices \cite{Giovannetti2007},
	\begin{equation}
		\mathcal{H}_{\Delta}= \Delta\bm{\sigma_z}\otimes\bm{s_0},
		\label{hd}
	\end{equation}
with $2\Delta$ denoting the magnitude of site energy difference between the two sublattices. When the staggered potential is present together with the intrinsic spin-orbit coupling, the electronic band structure becomes spin and valley dependent.

The staggered potential can also exist in free-standing systems with different atoms in the two sublattices (possessing the $D_{3h}$ point group symmetry), like for instance in hexagonal boron nitride (BN). Moreover, in such a case one can also introduce different intrinsic spin-orbit coupling parameters in the two sublattices, so instead of $\mathcal{H}_{so}^{int}$ (see equation \eref{hd6h}) one should use the following one
	\begin{equation}
		\mathcal{H}_{so}^{int}=\frac{\tau}{2}\left[\lambda_A(\bm{\sigma_z+\sigma_0})+\lambda_B(\bm{\sigma_z-\sigma_0}) \right]\otimes\bm{s_z}
		\label{hisoc},
	\end{equation}
with $\lambda_A$ and $\lambda_B$ denoting the spin-orbit parameters in the two sublattices and $\bm{\sigma_0}$ being $2\times 2$ identity matrix in the pseudospin space. The interaction given by equation \eref{hisoc} applies for instance to low-energy excitations near high symmetry points in BN, MoS$_2$, and generally in  group-VI dichalcogenide monolayers \cite{Xiao2012}.
Width of the band gap is given by the following formula: $\tau\zeta(\lambda_B+\lambda_A)+2\Delta$, where $\zeta =1(-1)$ for spin up (down).
	
Since our main interest is in the modification of tunnelling in junction of graphene-like materials which inevitably have to be deposited on a substrate, we focus mainly on substrate-induced interactions, and take into account the pseudospin dependent proximity exchange \cite{Zollner2016}:
	\begin{equation}
		\mathcal{H}_{ex}=\frac{1}{2}\left[\lambda_{ex}^A(\bm{\sigma_z+\sigma_0})+\lambda_{ex}^B(\bm{\sigma_z-\sigma_0}) \right]\otimes\bm{s_z},
		\label{hiex}
	\end{equation}
where $\lambda_{ex}^A$ and $\lambda_{ex}^B$ are exchange parameters for the sublattices A and B, respectively. Interestingly, this interaction has the form similar to the Hamiltonian \eref{hisoc}, except for the factor $\tau$ (i.e. it has the same sign for both Dirac points).

According to the above, the structures considered in this paper are well described by the Hamiltonian
	\begin{equation}
	\eqalign{
		\mathcal{H}&=\mathcal{H}_{0}+\mathcal{H}_{\Delta}+\mathcal{H}_{ex} = \\
		&=\left(\begin{array}{cc}
		\zeta\lambda_{ex}^A+\Delta & \hbar v_f (-\imath \tau \frac{\partial}{\partial x} + \frac{\partial}{\partial y})   \\
		\hbar v_f ( -\imath \tau \frac{\partial}{\partial x} - \frac{\partial}{\partial y}) & -\zeta\lambda_{ex}^B-\Delta \\
		\end{array} \right)
		}
	\label{hprox}
	\end{equation}	
for spin up ($\zeta=1$) and spin down ($\zeta=-1$).
The corresponding electronic band structure is then defined by the following dispersion relation,
	\begin{equation}
		\mathcal{E}_{\zeta}({\bf k})= \lambda_{ex}^- \pm \sqrt{\hbar^2 v_f^2(k_x^2+k_y^2)+\left(\lambda_{ex}^+ +\Delta \right)^2},
		\label{eprox}
	\end{equation}
with $\lambda_{ex}^\pm=\zeta(\lambda_{ex}^A\pm\lambda_{ex}^B)/2$. Band gap edges are defined by $(\zeta\lambda_{ex}^A+\Delta)$ and $-(\zeta\lambda_{ex}^B+\Delta)$.
	
	\subsection{Junctions of 2D materials}\label{Scenario}
	
	\indent \indent Let us consider now junctions based on the materials described by Hamiltonian \eref{hprox}. For theoretical calculations, we assume the barrier is rectangular. Without loss of generality, we assume the barrier of width D is along the axis $y$. Such a barrier can be formed by appropriate external gating of the system. We also assume the whole system is uniform in the $y$ direction, so the external gating potential (measured in energy units) can be written as:
	\begin{equation}
		V(x)=\cases{
				V& $0<x<D$\\
				0& otherwise.
				}
		\label{v}
	\end{equation}
The corresponding dispersion relation in the barrier region has the following form:
	\begin{equation}
		\mathcal{E}_{\zeta}^{V}({\bf q}) = V+\lambda_{ex}^- \pm \sqrt{\hbar^2 v_f^2(q_x^2+q_y^2)+\left(\lambda_{ex}^+ +\Delta \right)^2},
		\label{eproxv}
	\end{equation}
where $q_x$ and $q_y$ are the wavevector components in the barrier. In turn, the dispersion relation $\mathcal{E}_{\zeta}({\bf k})$ for external regions (left and right of the barrier) is given by equation \eref{eprox} (or equation \eref{eproxv} with $V=0$ and $q_x, q_y$ replaced by $k_x, k_y$).
	\begin{figure}
		\includegraphics[width=0.49\textwidth]{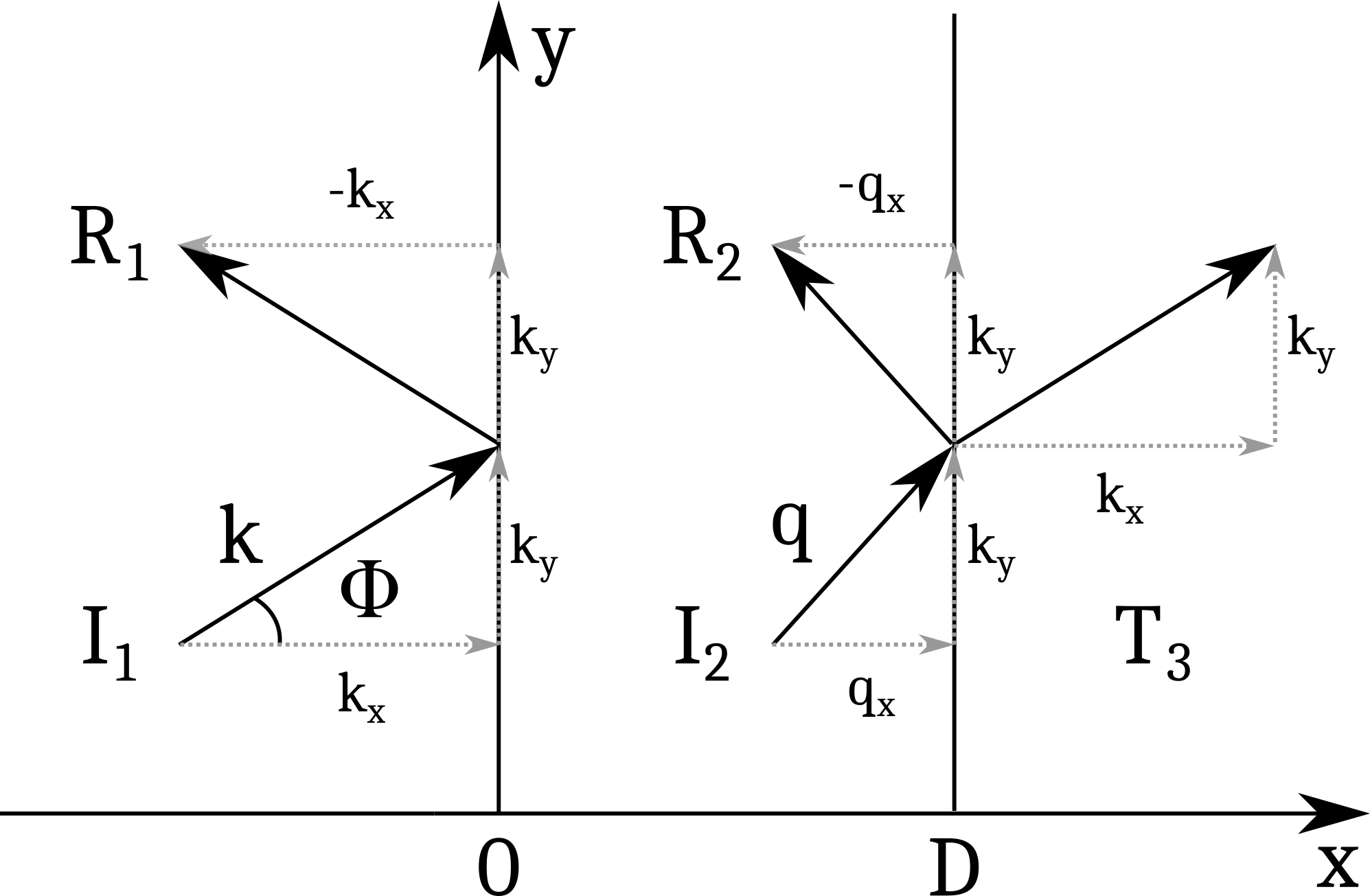}
		\captionof{figure}{Two dimensional wavevectors in the three regions of a rectangular potential barrier $V(x)$, with the incident angle marked by $\phi$.}
		\label{fig:wavevectors}
	\end{figure}
The electron waves, incident from the left, are partly reflected back and partly transmitted to the barrier region, and then transmitted to the right, as shown schematically
in figure \ref{fig:wavevectors}. The wavevector component along the barrier, ($k_y$ component) is conserved, $k_y=q_y=|{\bf k}|\sin \phi$, where $\phi$ is the incident angle (see figure \ref{fig:wavevectors}).

Outside the barrier region, eigenstates of the Hamiltonian \eref{hprox} for right and left propagating electron wave have the following form:
	\begin{equation}
		\eqalign{
			\Psi_R &=
			\left(\begin{array}{c}
			\mathcal{E}_{\zeta}+\zeta\lambda_{ex}^B+\Delta\\
			\hbar v_f(\tau k_x -\imath k_y)
			\end{array} \right)
			e^{\imath (k_x x +k_y y)},  \nonumber \\
			\Psi_L &=
			\left(\begin{array}{c}
			\mathcal{E}_{\zeta}+\zeta\lambda_{ex}^B+\Delta\\
			-\hbar v_f (\tau k_x+\imath k_y)\\
			\end{array} \right)
			e^{\imath (-k_x x +k_y y)}.
			}
		\label{eigenk}
	\end{equation}
For the notation simplicity, the spin index at the wave functions is suppressed here and will be also suppressed till the end of section 2.
The eigenstates $\Psi_{L(R)}^V$ in the barrier region are given by equation \eref{eigenk}  with the following replacements: $\mathcal{E}_{\zeta} \to \mathcal{E}_{\zeta}^V$ and $k_x \to q_x$, with
	\begin{equation}
		q_x=\sqrt{\frac{\left(E-V-\lambda_{ex}^-\right)^2-\left(\lambda_{ex}^+ +\Delta\right)^2}{\hbar^2 v_f^2}-k_y^2}.
		\label{qx}
	\end{equation}
Inside the barrier, the wavevector component $q_x$ becomes imaginary when the following condition is fulfilled:

	\begin{equation}
		\eqalign{
			\left(E-V-\lambda_{ex}^- \right)^2-\left(\lambda_{ex}^+ +\Delta\right)^2 + \\
			- \left[\left(E-\lambda_{ex}^-\right)^2-\left(\lambda_{ex}^+ + \Delta\right)^2 \right]\sin^2(\phi) < 0.
			}
		\label{imaginaryqx}
	\end{equation}

The entire ansatz for the considered transmission can be written in the following form:

	\begin{equation}
		\eqalign{
			\Psi_{1} = &\Psi_R+ R_1 \Psi_L, \\
			\Psi_{2} = I_2 &\Psi_R^V + R_2 \Psi_L^V, \\
			\Psi_{3} = T_3 &\Psi_R.
			}
		\label{ansatz}
	\end{equation}
After applying appropriate continuity conditions, a set of following probability amplitudes is obtained:
	\begin{equation}
		R_1=-\frac{( P + \imath\tau O^-) (P +
			\imath  \tau O^+) \sin(q_x D)}{\imath 2 k_x  q_x N \cos(q_x D) + Q \sin(q_x D)},
		\label{complexR1}
	\end{equation}
	\begin{equation}
		I_2=\frac{e^{-\imath q_x D} k_x (\tau P + \imath O^+)}{
			\imath 2 k_x q_x N \cos(q_x D) + Q\sin(q_x D)},
		\label{complexI2}
	\end{equation}
	\begin{equation}
		R_2=\frac{e^{\imath q_x D} k_x( \tau P + \imath O^-)}{-\imath 2 k_x q_x N \cos(q_x D) - Q \sin(q_x D)},
		\label{complexR2}
	\end{equation}
	\begin{equation}
		T_3=\frac{\imath 2 e^{-\imath k_x D} k_x q_x N}{\imath 2 k_x q_x N \cos(q_x D) + Q\sin(q_x D)},
		\label{complexT3}
	\end{equation}
where
\begin{equation}
		\eqalign{
			N&= \frac{E-V+\Delta+\zeta\lambda_{ex}^B}{E+\Delta+\zeta\lambda_{ex}^B}, \qquad	P=k_y(N-1), \\ O^{\pm}&=k_x N \pm q_x, \qquad Q=P^2 + k_x^2 N^2+ q_x^2.
			}
		\label{NPQ}
	\end{equation}
The corresponding transmission probability $t$ takes then the form:
	\begin{equation}
		\eqalign{
			t=T_3 T_3^*= \\
			=\left\{\cos^2(q_x D)+ \sin^2(q_x D)\left[\frac{P^2+N^2 k_x^2+q_x^2}{2 N k_x q_x}\right]^2\right\}^{-1},
			}
		\label{t}
	\end{equation}
which is applicable for real as well as imaginary $q_x$. Note, the probability amplitudes and the transmission probability $t$ depend generally on spin orientation but this dependence is not indicated here explicitly.

	\subsection{Bidirectional group delay}\label{Bidir}
	
	\indent \indent
As already described in the introduction, an important problem in tunnelling of particles (or quasi-particles) through a barrier is group delay and the associated Hartman effect. Following \cite{Winful2006}, we construct a wave packet, that is spatially localized with a Gaussian $f(E-E_0)$ distribution centred at $E_0$.
The wave packet upon interaction with the barrier becomes split into two wave packets, one reflected and one transmitted:
	\begin{equation}
		\eqalign{
			\Psi_{T_3}&=\int f(E-E_0)T_3(E)e^{\imath k_x x+\imath k_y y}e^{-\imath E t/\hbar}dE , \\
			\Psi_{R_1}&=\int f(E-E_0)R_1(E)e^{-\imath k_x x +\imath k_y y}e^{-\imath E t/\hbar}dE .
			}
		\label{wavepacketrt}
	\end{equation}
Amplitudes of transmission and reflection can be generally expressed as $|T_3(E)|e^{\imath\phi_t(E)}$ and $|R_1(E)|e^{\imath\phi_r(E)}$. For a sufficiently narrow distribution function, the magnitude of transmission and reflection coefficients can be considered constant over the integrating range: $|T_3(E)|\approxeq|T_3|$. This means wave packet is not reshaped and its group velocity (outside the barrier) along the x axis is 	
	\begin{equation}
		v_g=\frac{v_f \cos(\phi) \sqrt{(E- \lambda_{ex}^-)^2-(\lambda_{ex}^+ + \Delta)^2 }}{E - \lambda_{ex}^-}.
		\label{groupvhex}
	\end{equation}
Dividing the barrier width by the group velocity, one finds the time needed to traverse the free space distance equivalent to the barrier thickness, often referred to as the equal time $t_0$.
		
Utilizing the stationary phase method, and assuming peak of the incident wave packet was initially (time $t=0$) at $x=0$, one can calculate the time needed for the transmitted wave packet peak to appear at the
barrier end,
	\begin{equation}
		\hbar\frac{\partial}{\partial E}(\phi_t+k_x D)=\tau_{gt},
		\label{groupdelay}
	\end{equation}
where \eref{groupdelay} is the group delay of the transmitted wave packet. Similar expression for the reflected wave packet is $\hbar \partial/\partial E \phi_r=\tau_{gr}$. For potentials symmetric around zero, that is $V(x)=V(-x)$, both expression gives the same result: $\tau_{gt}=\tau_{gr}$. For other cases, it is useful to define bidirectional group delay, given by:
	\begin{equation}
		\tau_g=|T_3|^2\tau_{gt}+|R_1|^2\tau_{gr}.
		\label{bidirgroupdelay}
	\end{equation}	
The group delay should not be taken as the time needed to traverse the barrier by the incident wave packet. The wave packet ceases to exist as soon as the interaction with the potential begins. Moreover, it is impossible to locate the peak, as the wave packet has a certain spatial extension and therefore some of its parts will reflect sooner than the others, causing interference effects. In a strict sense, group delay $\tau_{gt}$ is the measure of time needed for the extrapolated peak of the transmitted wave packet to appear at $x=D$, with the assumption that extrapolated incident wave packet peak was at $x=0$ for $t=0$.
	
In certain conditions, the group delay becomes constant for sufficiently wide barriers. If the group delay is taken incorrectly as the traverse time of the distance equal to barrier width, then the following speed can be associated with the act of quantum tunnelling:
	
	\begin{equation}
		v=\frac{D}{\tau_g}
		\label{hartmanvg}
	\end{equation}
Since $\tau_g$ tends to a constant value, this means $v$ is increasing with the barrier thickness. Since there is no upper limit on the width of the barrier, $v$ can achieve superluminal speeds. Such property is often referred to as the Hartman effect.

	\section{Graphene on Co(Ni) with BN spacing}\label{ferro}

\begin{figure*}
        \begin{center}
            \includegraphics[width=1\textwidth]{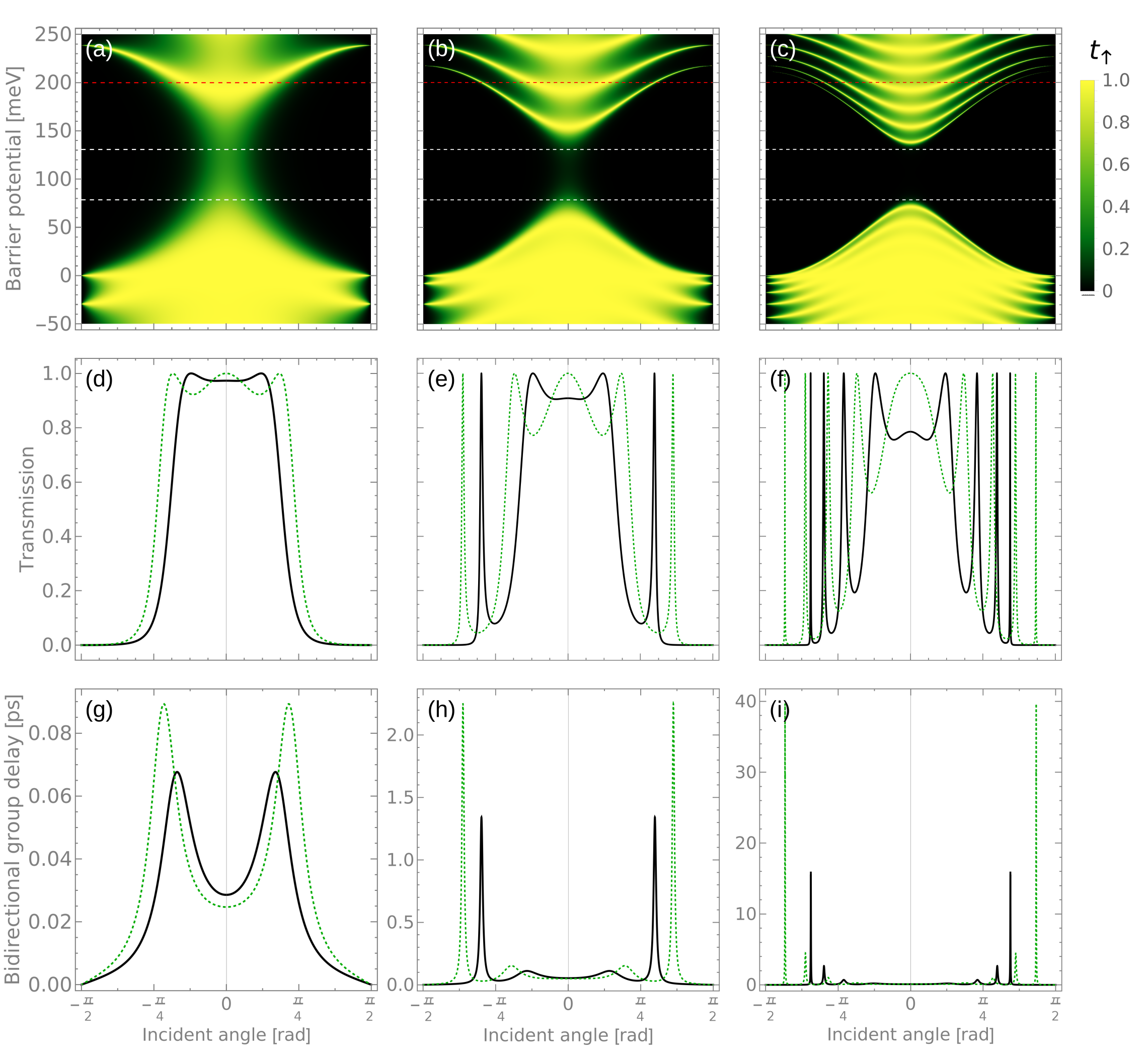}
        \end{center}
        \captionof{figure}{Transmission probability (a-f) and bidirectional group delay (g-i) for spin up in the G/BN(1)/Ni heterostructure as a function of the angle of incidence and barrier potential (a-c) for the energy of incident electrons equal to $100$ meV. The potential barrier in (d-i) is $V=200$ meV. Left column is	for the barrier width $D=20$ nm, middle column for $D=40$ nm,  and the right column for $D=80$ nm. The energy gap between the valence and conduction bands in the barrier is marked in (a-c) by the white dashed lines. The dotted green lines in (d-i) correspond to gapless graphene (described by the Hamiltonian \eref{h0}) and are included for comparison.}
        \label{fig:dwidthbandgap}
    \end{figure*}

\indent \indent Direct deposition of graphene onto a conducting ferromagnet changes band structure substantially due to hybridization with metallic orbitals, and such system does not exhibit Dirac cones. Moreover, transport is then not limited to graphene but takes place in the whole system. To avoid these undesired effects, the graphene layer is assumed to be separated from the metallic substrate by a few atomic planes of BN which is a wide gap insulator \cite{Ba2017} and also a good material for depositing graphene onto it \cite{Dean2010, Gannett2011}. The ferromagnetic metallic substrate induces then a uniform exchange field in the graphene monolayer. The proximity exchange  across BN exhibits properties reminiscent of the interlayer exchange coupling in magnetic metallic multilayers and especially in magnetic tunnel junctions. Not only magnitude but also sign of this proximity induced effect depends on the insulating layer thickness. However, the coupling almost disappears already for 3 BN monolayers.
Recently, Zollner \etal \cite{Zollner2016} performed ab-initio calculations and fitted the obtained dispersion relations near the Dirac points to the electronic spectrum derived from the Hamiltonian \eref{hprox}. \Tref{table:tab1} shows all the parameters obtained from such a fitting, which are relevant to our work.

\begin{figure}
	\includegraphics[width=0.44\textwidth]{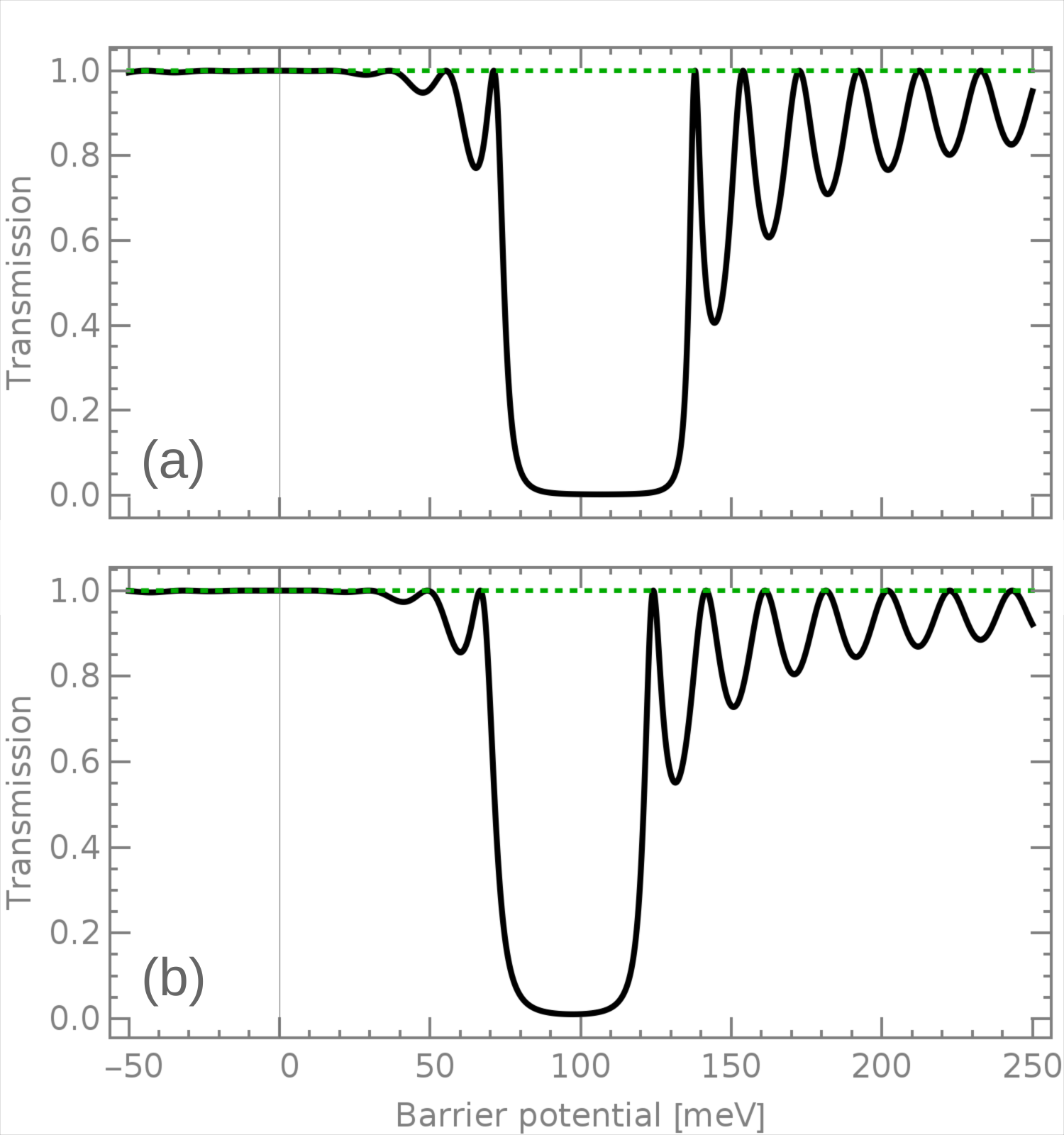}
	\captionof{figure}{Tunneling probability in G/BN(1)/Ni system for normal incidence as a function of the barrier height for spin up (a) and down (b). The energy of incident electrons equals $100$ meV, barrier width $D=80$ nm.}
	\label{fig:klein}
\end{figure}

	\begin{table}
	\caption{Parameters of effective Hamiltonian \eref{hprox} for  G/BN(i)/Co and G/BN(i)/Ni systems, and for i (i=1-3) monolayers of BN. Parameters from \cite{Zollner2016}.}
	\begin{indented}
		\lineup
		\item[]
		\begin{center}
			\begin{tabular}{@{}*{6}{c}}
				\br
				& $\lambda_{ex}^A$ & $\lambda_{ex}^B$ & $\Delta$ & $v_f$ \\
				& [meV]& [meV] & [meV] & [km/s]\\
				\mr
				G/BN(1)/Ni\ & \-1.4 & 7.78 & 22.86 &810\cr
				G/BN(2)/Ni \ & 0.068 & \-3.38 & 42.04 & 824\cr
				G/BN(3)/Ni \ & \-0.005 & 0.017 & 40.57 & 826\cr
				\mr
				G/BN(1)/Co \ & \-3.14 & 8.59 & 19.25 &812\cr
				G/BN(2)/Co \ & 0.097 & \-9.81 & 36.44 & 820\cr
				G/BN(3)/Co \ & \-0.005 & 0.018 & 38.96 & 821\cr
				\br
			\end{tabular}
		\end{center}
	\end{indented}
	\label{table:tab1}
\end{table}

\begin{figure}
        \includegraphics[width=0.43\textwidth]{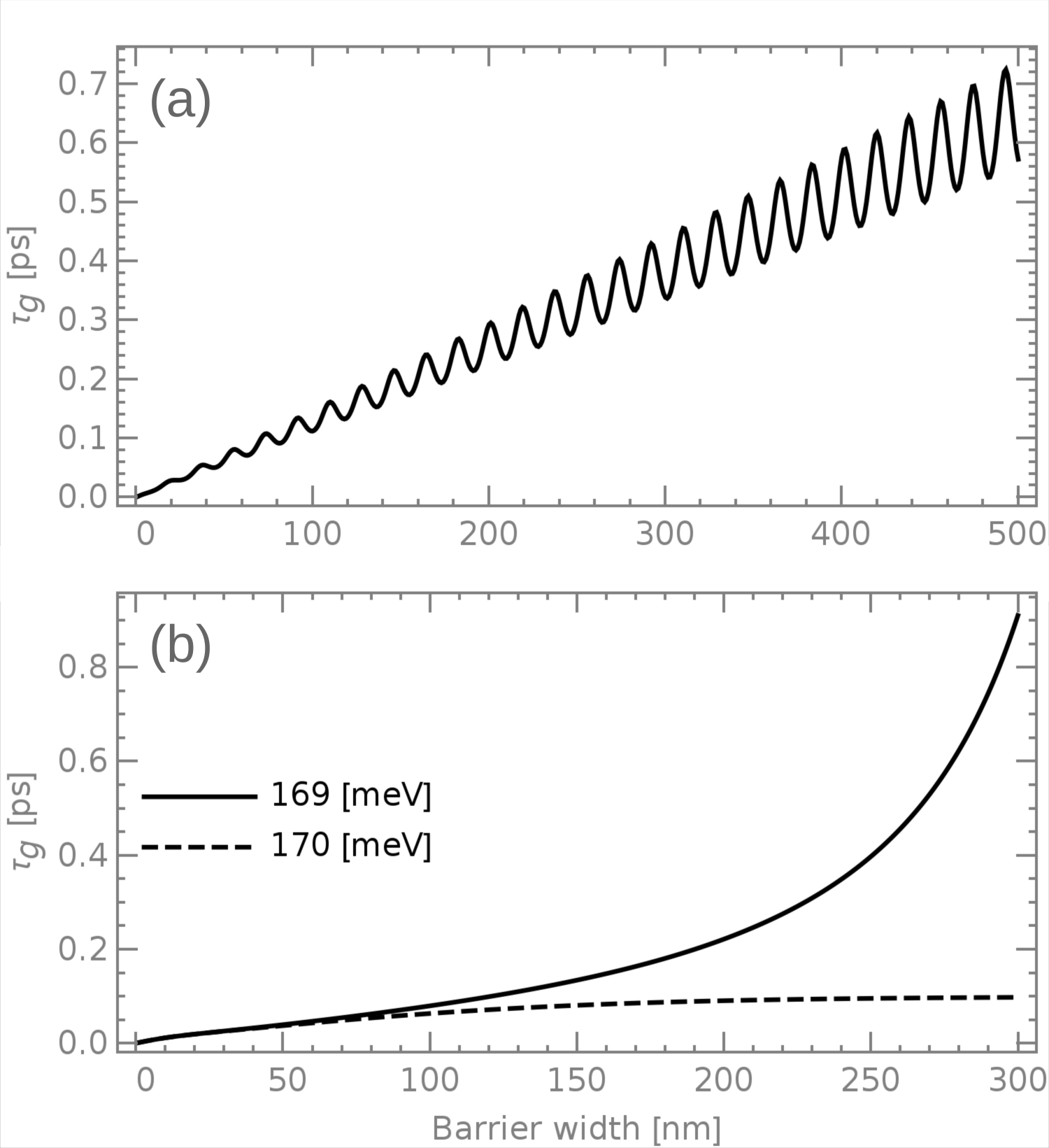}
    \captionof{figure}{Bidirectional group delay in G/BN(1)/Ni system for normal incidence of spin up electrons, $V=200$ meV, and energy of incident electrons $100$ meV (a) and $169$ and $170$ meV (b).}
        \label{fig:hartman}
\end{figure}
	
Though certain transport characteristics of junctions under consideration are similar to those of junctions made from pristine graphene, there are some qualitative and quantitative differences. These appear due to staggered potential and exchange coupling with the ferromagnetic substrate through BN. First of all, there is a gap in electronic structure between the conduction and valence bands. Second, owing to the proximity exchange interaction, tunnelling through the barrier becomes additionally spin-dependent. In figure \ref{fig:dwidthbandgap}(a-c) we show the variation of the tunnelling probability with the angle of incidence $\phi$ and the barrier height $V$ for spin up electrons in the G/BN(1)/Ni heterostructure and for three different values of the barrier width. The energy of incident electrons is assumed there to be equal to $100$ meV, therefore is within the conduction band.
Barrier potential effectively shifts band structure inside the barrier by $V$, relative to outer regions. For the barrier height between the white dashed lines in figure \ref{fig:dwidthbandgap}(a-c) the incident energy corresponds to the energy gap in the barrier.
The calculated parameters of the band gap for both spins are listed in \tref{table:tab2}.
Features due to the band gap are clearly seen in this figure. For small barrier heights, the transmission coefficient is large since there are available electronic states in the barrier for the assumed energy of tunnelling electrons. When the barrier height increases, the tunnelling electrons enter the band gap in the barrier and tunnelling probability is strongly reduced. It is worth to note that transmission probability behaves then similarly to that for free electrons governed by the Schr\"{o}dinger equation. When the barrier height increases further, the transmission probability becomes large again and approaches that for small barrier heights, as the energy of incident electrons is then within the valence band in the barrier.

For sufficiently wide barriers, the transmission probability for normal incidence is reduced to almost zero  when energy of the incoming quasiparticles is within the energy gap in the barrier. In the case shown in figure \ref{fig:dwidthbandgap} this appears already for barrier width of $80$ nm. The
band gap has two important consequences. First, tunnelling probability for normal incidence and for energy corresponding to propagative states in the barrier is generally reduced to values smaller than unity, as shown in figures \ref{fig:dwidthbandgap}(d-f), where the angular dependence of tunnelling probability is shown explicitly. The energy of incident electrons is assumed there to be $100$ meV, while the barrier potential is $V=200$ meV. These plots correspond to the red dashed lines in figures \ref{fig:dwidthbandgap}(a-c). For comparison, tunnelling probability in junctions made of pristine graphene is shown in figures \ref{fig:dwidthbandgap}(d-f) by the dotted green lines. In the latter case, the tunnelling probability is equal to 1 for normal incidence, which is the essence of the Klein
tunnelling. In the systems under considerations, full transmission appears for specific angles of incidence as shown in figures \ref{fig:dwidthbandgap}(d-f). However, from figures \ref{fig:dwidthbandgap}(a-c) one may conclude that
one can achieve
perfect transmission for normal incidence at some specific values of the barrier potential. Indeed, this is the case as shown in figure \ref{fig:klein}, where the tunnelling probability is shown as a function of the barrier potential. The
transmission drops to almost zero when tunnelling is through the band gap in the barrier and oscillates with the potential barrier when tunnelling is due to propagative states in the barrier. The tunnelling probability reaches then unity for certain values of the barrier potential. This behaviour is in contrast with pristine graphene, where the Klein tunnelling for normal incidence appears independently of the barrier potential (see the green dotted lines in figure \ref{fig:klein}).

Another phenomenon associated with tunnelling is the Hartman effect which is related to the group delay. Angular dependence of the bidirectional group delay (see equation \eref{bidirgroupdelay}) is shown in  figures \ref{fig:dwidthbandgap}(g-i) for three barrier widths. These figures correspond to the red dashed lines in figures \ref{fig:dwidthbandgap}(a-c). First, maxima of the group delay coincide with the maxima in tunnelling probability. The group delay is especially low in the areas of small tunnelling probability. Second, the maxima increase with increasing barrier width.
This  is shown explicitly in figure \ref{fig:hartman} for normal incidence, where the bidirectional group delay is presented as a function of the barrier width. Figure \ref{fig:hartman}(a) is for the incident energy equal to $100$ meV, which corresponds to propagative modes in the barrier. The group delay is then small and increases on average linearly with the barrier width, revealing weak oscillation around the average value. In figure \ref{fig:hartman}(b) the two curves correspond to the  electron energy $169$ meV and $170$ meV. For the structure G/BN(1)/Ni and spin up orientation, the lowest energy needed to induce imaginary wavevector inside the barrier is above $169.36$ meV \eref{imaginaryqx}. Accordingly, the curve for $170$ meV corresponds to tunnelling due to evanescent waves and the related group delay clearly saturates with increasing barrier width, which is the evidence of the Hartman effect.

	\begin{table}
		\caption{The band gap parameters of G/BN(i)/Co and G/BN(i)/Ni systems for $i=1$-$3$. Here, $Eg$ denotes gap width, $L$ is the lower gap edge, while $H$ the upper band gap edge. The subscripts $\uparrow$ and $\downarrow$ correspond to spin up and spin down orientation, respectively. All values in meV.}
		\begin{indented}
			\lineup
			\item[]
			\begin{center}
				\begin{tabular}{@{}*{7}{c}}
					\br
					& $L_{\uparrow}$ & $L_{\downarrow}$ & $H_{\uparrow}$ & $H_{\downarrow}$ & $Eg_{\uparrow}$ & $Eg_{\downarrow}$ \\
					\mr
					G/BN(1)/Ni & \-30.64 & \-15.08 & 21.46 & 24.26 & 52.1  & 39.34 \cr
					G/BN(2)/Ni & \-38.66 & \-45.42 & 42.11 & 41.97 & 80.77 & 87.39 \cr
					G/BN(3)/Ni & \-40.59 & \-40.55 & 40.56 & 40.57 & 81.15 & 81.13 \cr
					\mr
					G/BN(1)/Co & \-27.84 & \-10.66 & 16.11 & 22.39 & 43.95 & 33.05 \cr
					G/BN(2)/Co & \-26.63 & \-46.25 & 36.54 & 36.34 & 63.17 & 82.59 \cr
					G/BN(3)/Co & \-38.98 & \-38.94 & 38.95 & 38.96 & 77.93 & 77.91 \cr
					\br
				\end{tabular}
			\end{center}
		\end{indented}
		\label{table:tab2}
	\end{table}

\begin{figure}
	\begin{center}
		\includegraphics[width=0.5\textwidth]{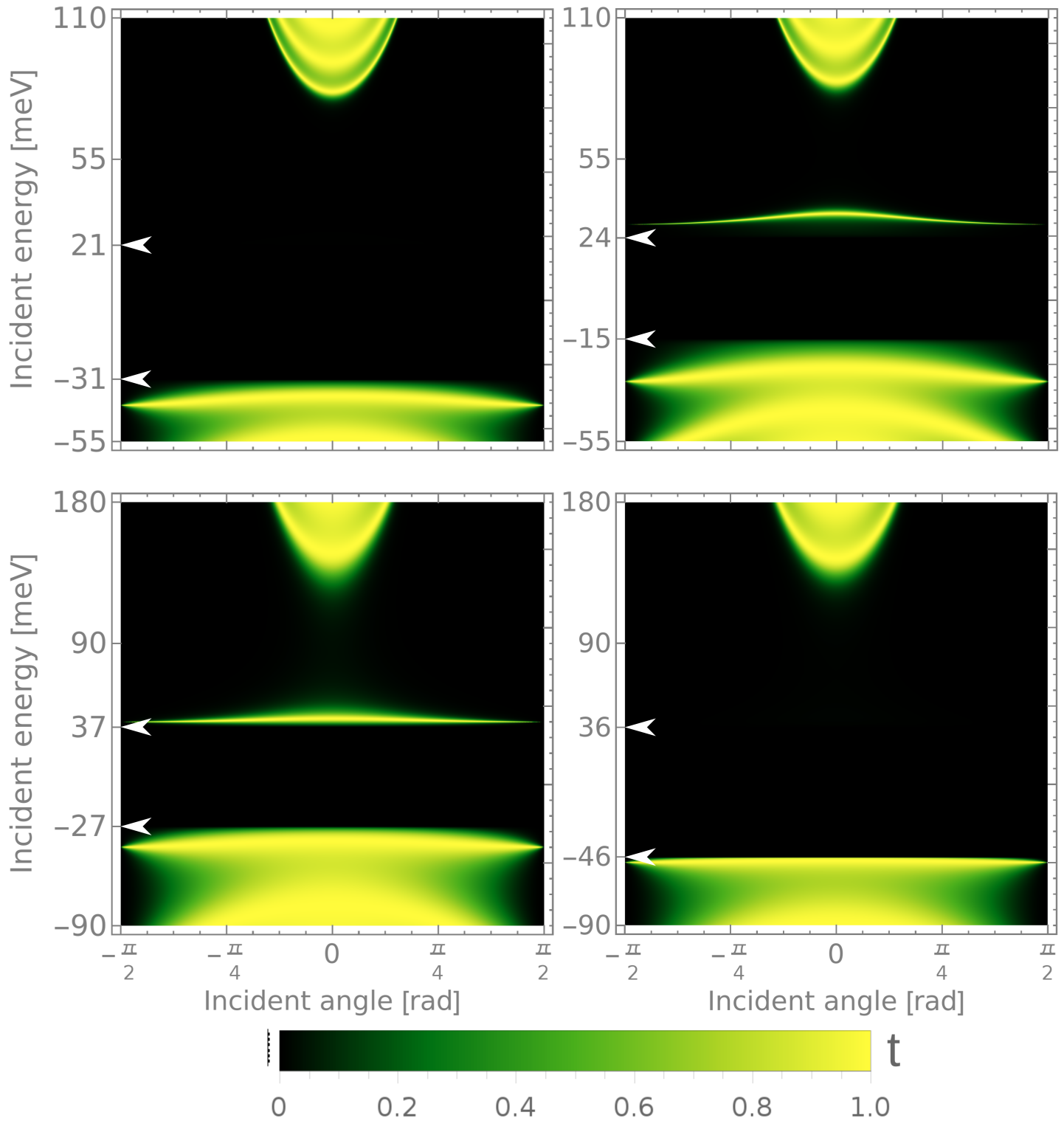}
	\end{center}
	\captionof{figure}{Spin-polarized tunnelling. The left(right) column is for spin up(down) electrons. The upper row is for G/BN(1)/Ni, $V=55$ meV, and $D=100$ nm. The lower row is for G/BN(2)/Co, $V=90$ meV, and $D=40$ nm. Band gaps are marked by arrowheads.}
	\label{fig:spinpol}
\end{figure}
	
Interesting from the experimental point of view is the situation when tunnelling for one spin orientation is strongly suppressed while tunnelling for the opposite spin is quite significant. This might take place when the upper edge of the gap in the electrodes coincides with the lower edge of the  gap in the barrier region for one spin orientation only. Upon careful tuning of the parameters, this presents a chance of fabricating a setup exhibiting fully spin-polarized electronic transport. Such systems may be used as efficient spin filters. The proposed parameters for G/BN(1)/Ni and G/BN(2)/Co heterostructuresb are presented in figure \ref{fig:spinpol}. For the nickel substrate, spin down quasiparticles with energy $\approx30$ meV
($\approx-500$ meV including the predicted in \cite{Zollner2016} Dirac
energy shift) are expected to be transmitted, while tunnelling of spin up electrons is expected to be fully suppressed in a wide range of energies around this particular value. For the cobalt substrate, spin up quasiparticles with energy $\approx40$ meV ($\approx-310$ meV) should tunnel while spin down particles within a large range of nearby energies should not.

	\section{Graphene on ferromagnetic insulators}\label{insulators}
	
\indent \indent Another approach aimed at inducing spin-polarized band structure in graphene 
is based on depositing graphene directly on a ferromagnetic insulator. This insulator induces a uniform exchange field in graphene. To investigate tunnelling properties in such heterostructures we used the parameters for graphene deposited on EuS, EuO and YIG (Y$_3$Fe$_5$O$_{12}$), obtained by Hallal \etal in \cite{Hallal2017} from first principles calculations. In order to be consistent within our work, the fitting parameters have been converted to the form used by Zollner \etal in \cite{Zollner2016}. These parameters are given in \tref{table:tab3}. In turn, \tref{table:tab4} contains the band gap parameters for graphene on ferromagnetic insulators. We use spin-dependent values of Fermi velocities $v_{\uparrow}$ and $v_{\downarrow}$, fitted to dispersion relations by Song \cite{Song2018}.

All the examined ferromagnetic insulators used as substrates generate spin splitting of the band structure, which is larger than that induced by metallic Ni or Co {\it via} the BN spacer. However, due to the low Curie temperature of $\approx 69$ K for G/EuO \cite{Swartz2012} and even lower for G/EuS $\approx 16$ K \cite{Wei2016}, real applications of such heterostructures are expected to be extremely limited. On the other hand, heterostructures including YIG as a substrate (with bulk YIG Curie temperature of $\approx 550$ K) have been confirmed to reveal proximity induced exchange coupling at room temperatures \cite{Leutenantsmeyer2016}. Because of this, a more thorough analysis will be carried out mainly for the system consisting of graphene deposited on YIG.

	\begin{table}
	\caption{Parameters of effective Hamiltonian \eref{hprox} for graphene on ferromagnetic insulators, as converted from \cite{Hallal2017} and \cite{Song2018}.}
	\begin{indented}
		\lineup
		\item[]
		\begin{center}
			\begin{tabular}{@{}*{7}{lcccccc}}
				\br
				& $\lambda_{ex}^A$ & $\lambda_{ex}^B$ & $\Delta$ & $v_{\uparrow}$ & $v_{\downarrow}$ \\
				& [meV]& [meV] & [meV] & [km/s] & [km/s]\\
				\mr
				G/EuO(6) \ & 42 & \-24 & 58 & 1151 & 1400 \cr
				G/EuS(6) \ & 11.5 & 5 & 88 & 1203 & 1460 \cr
				G/Y$_3$Fe$_5$O$_{12}$ \ & \-26 & 57.5 & 42 & 541 & 601 \cr
				\br
			\end{tabular}
		\end{center}
	\end{indented}
	\label{table:tab3}
\end{table}

\begin{table}
	\caption{The band gap of graphene on ferromagnetic insulators. Here, $Eg$ denotes gap width, $L$ is the lower gap edge, while $H$ the upper band gap edge. The subscripts $\uparrow$ and $\downarrow$ correspond to spin up and spin down orientation, respectively. All values in meV.}
	\begin{indented}
		\lineup
		\item[]
		\begin{center}
			\begin{tabular}{@{}*{7}{lcccccc}}
				\br
				& $L_{\uparrow}$ & $L_{\downarrow}$ & $H_{\uparrow}$ & $H_{\downarrow}$ & $Eg_{\uparrow}$ & $Eg_{\downarrow}$ \\
				\mr
				G/EuO(6) & \-34 & \-82 & 100 & 16 & 134 & 98 \cr
				G/EuS(6) & \-93 & \-83 & 99.5  & 76.5 & 192.5 & 159.5 \cr
				G/Y$_3$Fe$_5$O$_{12}$ & \-99.5 & 15.5 & 16 & 68.5 & 115.5 & 52.5 \cr
				\br
			\end{tabular}
		\end{center}
	\end{indented}
	\label{table:tab4}
\end{table}

	\begin{figure*}
	\begin{center}
		\includegraphics[width=1\textwidth]{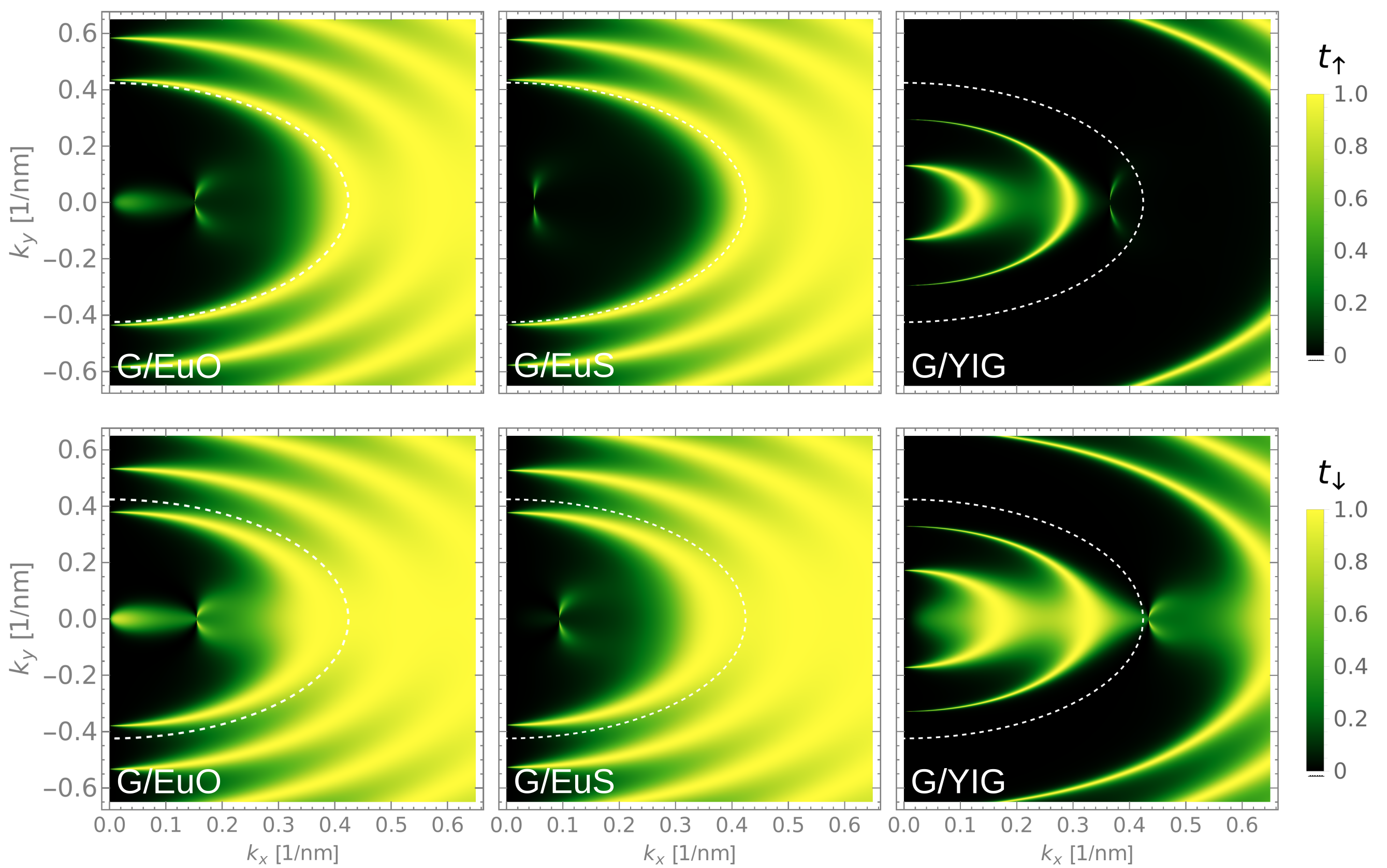}
	\end{center}
	\captionof{figure}{Transmission probability for positive incident energy. The upper row is for spin up, while lower one for spin down. Left column corresponds to graphene on EuO, middle to graphene on EuS, and the right one to graphene on YIG. The other parameters are: barrier width $D=20$ nm, and barrier potential $V=200$ meV. The semicircles of radius $\sqrt{0.18}$ 1/nm, marked by the white dashed lines, are added for reference.}
	\label{fig:tkupdown}
\end{figure*}

\begin{figure*}
	\begin{center}
		\includegraphics[width=1\textwidth]{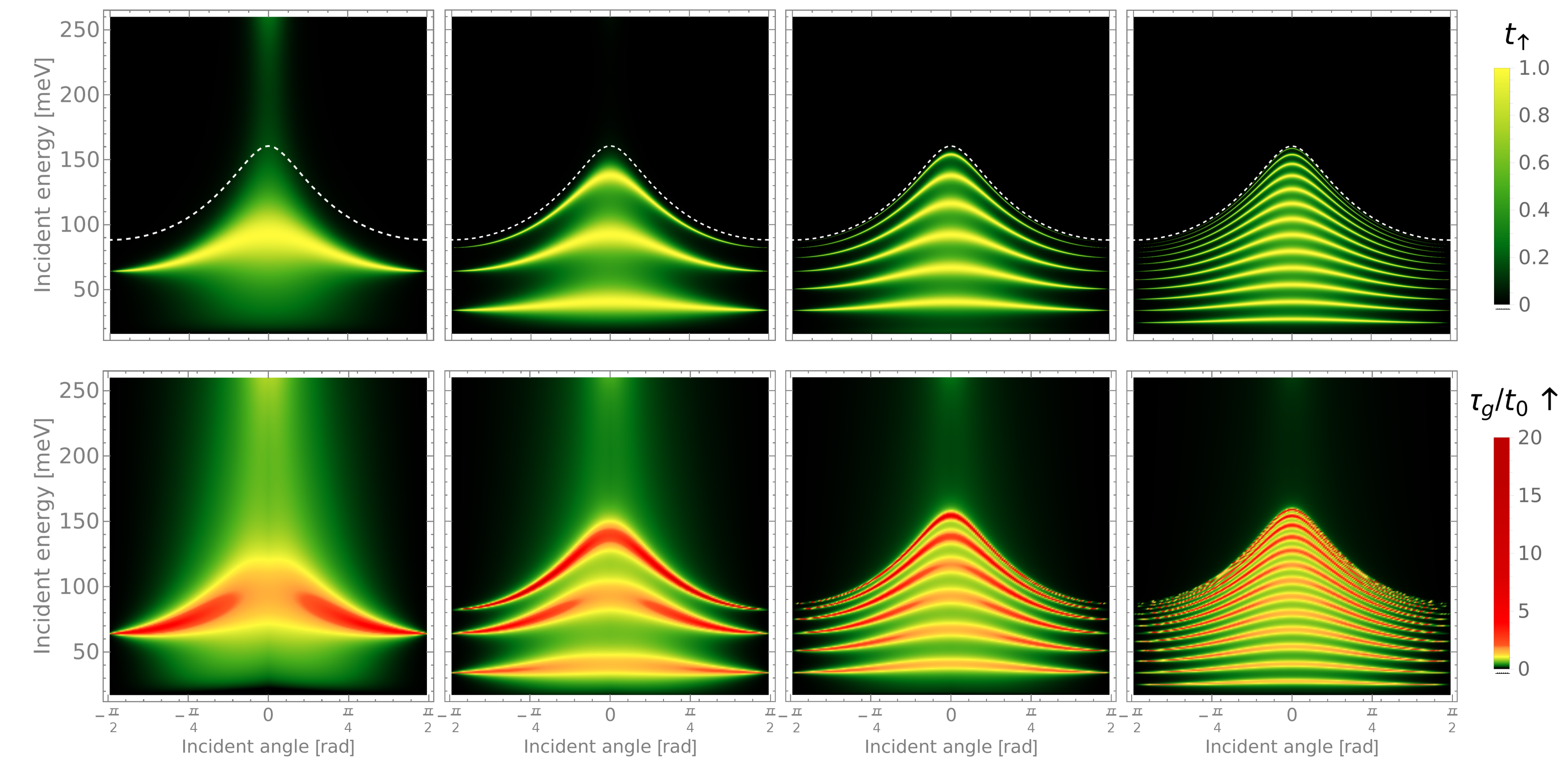}
	\end{center}
	\captionof{figure}{Transmission probability (upper row) and bidirectional group delay/$t_0$ (lower row) for spin up in graphene on YIG, and for $V=260$ meV. Columns correspond to the following barrier width: from the left $10$, $20$, $40$ and $80$ nm. The areas above white dashed lines in the top row correspond to imaginary $q_x$.}
	\label{fig:hartmanyig}
\end{figure*}

\begin{figure*}
	\begin{center}
		\includegraphics[width=1\textwidth]{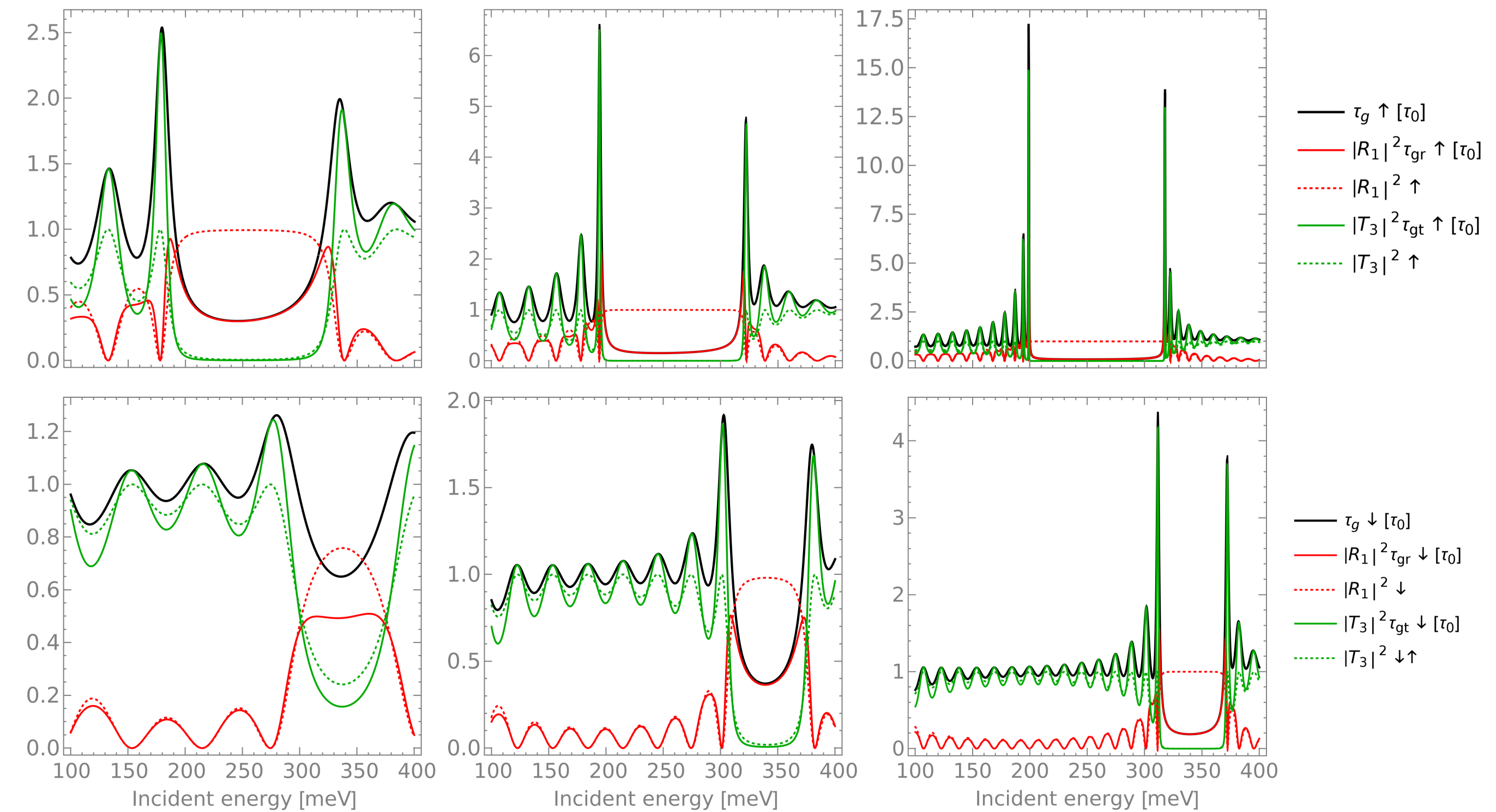}
	\end{center}
	\captionof{figure}{Bidirectional group delay for graphene on YIG for $V=300$ meV, $\phi=0$,  spin up (upper row) and spin down (lower row); and for the barrier widths, from the left: $20$, $40$ and $80$ nm.}
	\label{fig:hartmanyigupdown}
\end{figure*}

Figure \ref{fig:tkupdown} shows transmission probability as a function of the incident wavevector components ($k_x$, $k_y$) for graphene on the three investigated ferromagnetic insulators and for positive energy of incident electrons. The upper row corresponds to spin up and the lower one to spin down orientations. Due to exchange interaction with the ferromagnetic substrates, tunnelling in all systems is spin-dependent. For $k_y=0$ this figure corresponds to normal incidence while for $k_x\to 0$ the incidence angle tends to $\pm \pi /2$. Since the energy spectra in the systems under consideration are isotropic in the $k_x, k_y$ space, the semicircles of radius $k=\sqrt{ k_x^2 +k_y^2}$, centred at $k_x=k_y=0$, describe angular dependence of the transmission probability for the corresponding energy of incident electrons. White dashed lines in figure \ref{fig:tkupdown} mark semicircles of the radius $\sqrt{0.18}$ 1/nm for reference. This angular dependence, together with dependence on the energy of incident electrons, is shown explicitly in the figure \ref{fig:hartmanyig}. The upper row shows the transmission probability for spin up electrons tunnelling in junctions based on graphene deposited on YIG, while the lower row presents the ratio of the group delay to equal time $t_0$. Since the barrier height is assumed relatively large as compared to the energy range of incident electrons, only tunnelling transport through the valence band (and partly gap) in the barrier is visible there. Different columns in these two rows correspond to different barrier widths (increasing from left to right). As in the preceding section, maxima in the group delay are correlated with the maxima in the transmission probability. Since the angular dependence of the transmission probability and of the group delay is qualitatively similar to that found in graphene on BN and metallic ferromagnet, we will not discuss them in more details.

In turn, figure \ref{fig:hartmanyigupdown} shows the ratio of bidirectional group delay to equal time for both spin orientations, along with the components corresponding to the transmitted and reflected parts of the wave function. Excluding energies corresponding to tunnelling {\it via} evanescent waves (Hartman effect regime), $|T_3|^2\tau_{gt}$ provides the dominant contribution. Bidirectional group delay oscillates around equal time, aside from the specific case when energy of the incident quasiparticle approaches the band gap inside the barrier. In this region, phase of the transmission amplitude varies remarkably with energy and sharp peaks of the group delay can be seen. Similar, yet lower peaks can be seen for incident energies corresponding to just above the band gap in the barrier. Bidirectional group delay does not depend significantly on the potential value but rather on the band gap size and barrier width, with smaller band gaps and narrow barriers inducing lower delays. The spin dependence of the parameters under discussion is closely correlated with the spin dependence of the electronic structure induced by the exchange field due to ferromagnetic substrate. Its influence is clearly visible in figure \ref{fig:hartmanyigupdown} (compare upper and lower rows).

\section{Summary}

\indent \indent In this paper, we have studied the influence of exchange fields generated by ferromagnetic substrates on tunnelling probability in graphene junctions. Two classes of systems were analysed. In the first class, graphene was separated from the metallic ferromagnet by one, two, or three atomic planes of insulating hexagonal boron nitride. In the second class, graphene was deposited directly on an insulating ferromagnet. In both cases, electronic transport can be limited to graphene.

The main interest was tunnelling probability and its dependence on the angle of incidence, barrier height and width, and on the energy of incident electrons. This was analysed in the context of Klein tunnelling and Hartman effect. It has been shown that when transmission is due to propagative states, the tunnelling probability for normal incidence is generally reduced below unity, and thus the main trait of Klein tunnelling for pristine graphene quasiparticles is suppressed. Perfect transmission occurs only for specific values of the parameters (incident energy, barrier height, barrier width). In turn, the Hartman effect appears in the regions where tunnelling is {\it via} evanescent states in the barrier. We have also found that such systems can be effectively used as spin filters.


\section*{References}
\begin{thebibliography}{99}
	
	\bibitem{Merzbacher2002}
	Merzbacher, E.
	\textit{The Early History of Quantum Tunneling}
	Physics Today 55, 44 (2002).
	
	\bibitem{Esaki1974}
	Esaki, L.
	\textit{Long journey into tunneling}
	Proceedings of the IEEE 62, 825–831 (1974)
	
	\bibitem{MacColl1932}
	MacColl L A
	\textit{Note on the Transmission and Reflection of Wave Packets by Potential Barriers}
	Physical Review 40 621–6 (1932)
	
	\bibitem{Hartman1962}
	Hartman T E
	\textit{Tunneling of a Wave Packet}
	Journal of Applied Physics 33 3427–33 (1962)
	
	\bibitem{Winful2002}
	Winful H
	\textit{Energy storage in superluminal barrier tunneling: Origin of the Hartman effect}
	Optics Express 10 1491 (2002)
	
	\bibitem{Winful2006}
	Winful, H. G.
	\textit{Tunneling time, the Hartman effect, and superluminality: A proposed resolution of an old paradox}
	Physics Reports 436, 1–69 (2006)

    \bibitem{Wang2004}
    Wang, L.-G., Xu, J.-P., Zhu., S.-Y.
    {Negative Hartman effect in one-dimensional photonic crystal with negative refractive materials}
    Phys.Rev.E 70, 066624 (2004)

    \bibitem{Klos2018}
    K\l os, W., Dadoenkova, Y. S., Rych\l y, J., Dadoenkova, N. N., Lyubchanskii, I. L., Barna\'s, J.
    {Hartman effect for spin waves in exchange regime}
    Scientifc Reports, to be published


    \bibitem{Wu2009}
    Wu, Z., Chang, K., Liu, J. T., Li, X. J., Chan, K. S.
    {The Hartman effect in graphene}
    J. Appl. Phys. 105, 043702 (2009)
	
    \bibitem{Klein1929}
	Klein O
	\textit{Die Reflexion von Elektronen an einem Potentialsprung nach der relativistischen Dynamik von Dirac}
	Zeitschrift für Physik 53 157–65 (1929)

	
	\bibitem{Calogeracos1999}
	Calogeracos A, Dombey N
	\textit{Klein tunnelling and the Klein paradox}
	International Journal of Modern Physics A 14 631–43 (1999)
	
	
	\bibitem{Novoselov2004}
	Novoselov, K. S. et al.
	\textit{Electric Field Effect in Atomically Thin Carbon Films}
	Science 306, 666–669 (2004)
	
	\bibitem{Katsnelson2006}
	Katsnelson, M. I., Novoselov, K. S., Geim, A. K.
	\textit{Chiral tunnelling and the Klein paradox in graphene}
	Nature Physics 2, 620–625 (2006).
	
	\bibitem{Huard2007}
	Huard B, Sulpizio J A, Stander N, Todd K, Yang B, Goldhaber-Gordon D
	\textit{Transport Measurements Across a Tunable Potential Barrier in Graphene}
	Physical Review Letters 98 (2007)
	
	\bibitem{Gorbachev2008}
	Gorbachev R V, Mayorov A S, Savchenko A K, Horsell D W, Guinea F
	\textit{Conductance of p-n-p graphene structures with “air-bridge” top gates}
	Nano Letters 8 1995–9 (2008)
	
	\bibitem{Young2009}
	Young A F, Kim P
	\textit{Quantum interference and Klein tunnelling in graphene heterojunctions}
	Nature Physics 5 222–6 (2009)
	
	\bibitem{Schwierz2010}
	Schwierz F
	\textit{Graphene transistors}
	Nature Nanotechnology 5 487–96 (2010)
	
	\bibitem{Xu2018}
	Xu X, Liu C, Sun Z, Cao T, Zhang Z, Wang E, Liu Z and Liu K
	\textit{Interfacial engineering in graphene bandgap}
	Chemical Society Reviews 47 3059–99 (2018)
	
	\bibitem{Kawasaki2002}
	Kawasaki T, Ichimura T, Kishimoto H, Akbar A A, Ogawa T and Oshima C
	\textit{Double atomic layers of graphene/monolayer h-BN on Ni(111) studied by scanning tunneling microscopy and scanning tunneling spectroscopy}
	Surface Review and Letters 09 1459–64 (2002)
	
	\bibitem{Geim2013}
	Geim A K and Grigorieva I V
	\textit{Van der Waals heterostructures}
	Nature 499 419–25 (2013)
	
	\bibitem{Du2016}
	Du A
	\textit{In silico engineering of graphene-based van der Waals heterostructured nanohybrids for electronics and energy applications: Graphene-based van der Waals heterostructured nanohybrids}
	Wiley Interdisciplinary Reviews: Computational Molecular Science 6 551–70 (2016)
	
	\bibitem{Qian2015}
	Qian X, Wang Y, Li W, Lu J and Li J
	\textit{Modelling of stacked 2D materials and devices}
	2D Materials 2 032003 (2015)
	
	\bibitem{Han2014}
	Han W, Kawakami R K, Gmitra M and Fabian J
	\textit{Graphene spintronics}
	Nature Nanotechnology 9 794–807 (2014)
	
	\bibitem{Yang2013}
	Yang H X, Hallal A, Terrade D, Waintal X, Roche S and Chshiev M
	\textit{Proximity Effects Induced in Graphene by Magnetic Insulators: First-Principles Calculations on Spin Filtering and Exchange-Splitting Gaps}
	Physical Review Letters 110 (2013)
	
	\bibitem{Hallal2017}
	Hallal A, Ibrahim F, Yang H, Roche S and Chshiev M
	\textit{Tailoring magnetic insulator proximity effects in graphene: first-principles calculations}
	2D Materials 4 025074 (2017)
	
	\bibitem{Zollner2016}
	Zollner K, Gmitra M, Frank T and Fabian J
	\textit{Theory of proximity-induced exchange coupling in graphene on hBN/(Co, Ni)}
	Physical Review B 94 (2016)
	
	\bibitem{Swartz2012}
	Swartz A G, Odenthal P M, Hao Y, Ruoff R S and Kawakami R K
	\textit{Integration of the Ferromagnetic Insulator EuO onto Graphene}
	ACS Nano 6 10063–9 (2012)
	
	\bibitem{Wang2015}
	Wang Z, Tang C, Sachs R, Barlas Y and Shi J
	\textit{Proximity-Induced Ferromagnetism in Graphene Revealed by the Anomalous Hall Effect}
	Physical Review Letters 114 (2015)
	
	\bibitem{Leutenantsmeyer2016}
	Leutenantsmeyer J C, Kaverzin A A, Wojtaszek M and van Wees B J
	\textit{Proximity induced room temperature ferromagnetism in graphene probed with spin currents}
	2D Materials 4 014001 (2016)
	
	\bibitem{Wei2016}
	Wei P, Lee S, Lemaitre F, Pinel L, Cutaia D, Cha W, Katmis F, Zhu Y, Heiman D, Hone J, Moodera J S and Chen C-T
	\textit{Strong interfacial exchange field in the graphene/EuS heterostructure}
	Nature Materials 15 711–6 (2016)
	
	\bibitem{Dyrdal2017}
	Dyrda\l~A and Barna\'s J
	\textit{Anomalous, spin, and valley Hall effects in graphene deposited on ferromagnetic substrates}
	2D Materials 4 034003 (2017)
	
	\bibitem{Wallace1947}
	Wallece P R
	\textit{The Band Theory of Graphite}
	Physical Review 71 622–34 (1947)
	
	\bibitem{Kochan2017}
	Kochan D, Irmer S and Fabian J
	\textit{Model spin-orbit coupling Hamiltonians for graphene systems}
	Physical Review B 95 (2017)
	
	\bibitem{Jiang2017}
	Jiang L, Marconcini P, Hossian M S, Qiu W, Evans R, Macucci M and Skafidas E
	\textit{A tight binding and $\overrightarrow{{\boldsymbol{k}}}\cdot \overrightarrow{{\boldsymbol{p}}}$ study of monolayer stanene}
	Scientific Reports 7 (2017)
	
	\bibitem{Liu2014}
	Liu C-C, Guan S, Song Z, Yang S A, Yang J and Yao Y
	\textit{Low-energy effective Hamiltonian for giant-gap quantum spin Hall insulators in honeycomb X -hydride/halide ( X = N – Bi ) monolayers}
	Physical Review B 90 (2014)
	
	\bibitem{Giovannetti2007}
	Giovannetti G, Khomyakov P A, Brocks G, Kelly P J and van den Brink J
	\textit{Substrate-induced band gap in graphene on hexagonal boron nitride: Ab initio density functional calculations}
	Physical Review B 76 (2007)
	
	\bibitem{Xiao2012}
	Xiao D, Liu G-B, Feng W, Xu X and Yao W
	\textit{Coupled Spin and Valley Physics in Monolayers of MoS 2 and Other Group-VI Dichalcogenides}
	Physical Review Letters 108 (2012)
	
	\bibitem{Ba2017}
	Ba K, Jiang W, Cheng J, Bao J, Xuan N, Sun Y, Liu B, Xie A, Wu S and Sun Z
	\textit{Chemical and Bandgap Engineering in Monolayer Hexagonal Boron Nitride}
	Scientific Reports 7 (2017)
	
	\bibitem{Dean2010}
	Dean C R, Young A F, Meric I, Lee C, Wang L, Sorgenfrei S, Watanabe K, Taniguchi T, Kim P, Shepard K L and Hone J
	\textit{Boron nitride substrates for high-quality graphene electronics}
	Nature Nanotechnology 5 722–6 (2010)
	
	\bibitem{Gannett2011}
	Gannett W, Regan W, Watanabe K, Taniguchi T, Crommie M F and Zettl A
	\textit{Boron nitride substrates for high mobility chemical vapor deposited graphene}
	Applied Physics Letters 98 242105 (2011)
	
	\bibitem{Song2018}
	Song Y
	\textit{Electric-field-induced extremely large change in resistance in graphene ferromagnets} Journal of Physics D: Applied Physics 51 025002 (2018)
	
\end{thebibliography}
	\end{document}